\documentclass[conference,9pt]{IEEEtran}

\usepackage{mathptmx} % This is Times font

\usepackage{fancyhdr}
\usepackage[normalem]{ulem}
\usepackage[sort,compress,noadjust]{cite}
\usepackage{flushend}
\usepackage{enumitem,kantlipsum}
\usepackage{algorithm}
\usepackage{algpseudocode}
\usepackage{comment}
\usepackage{multirow}
\usepackage{xspace}
\usepackage{listings}
\usepackage{tikz}
\usepackage{xcolor}
\usepackage{pifont}
\usepackage{dblfloatfix}

\usepackage[bookmarks=true,breaklinks=true,colorlinks,linkcolor=black,citecolor=blue,urlcolor=black]{hyperref}

\def\Snospace~{\S{}}

\newcommand{\cc}{\texttt{crash-consistency}\xspace}
\newcommand{\sync}{\texttt{synchronization}\xspace}
\newcommand{\cs}{\texttt{crash-sync-safe}\xspace}
\newcommand{\csty}{\texttt{crash-sync-safety}\xspace}

\newcommand{\sys}{\texttt{ccHTM}\xspace}
\newcommand{\cchtm}{\texttt{ccHTM}\xspace}
\newcommand{\htm}{\texttt{HTM}\xspace}
\newcommand{\htmns}{\texttt{HTM}}  % \htm no space
\newcommand{\clflush}{\texttt{clflush}\xspace}
\newcommand{\clflushopt}{\texttt{clflushopt}\xspace}
\newcommand{\clwb}{\texttt{clwb}\xspace}
\newcommand{\sfence}{\texttt{sfence}\xspace}
\newcommand{\fence}{\texttt{fence}\xspace}

\newcommand{\ccms}{crash-consistency mechanisms\xspace}
\newcommand{\seq}{\texttt{seq}\xspace}
\newcommand{\undo}{\texttt{undo}\xspace}
\newcommand{\redo}{\texttt{redo}\xspace}
\newcommand{\tsxseq}{\texttt{\htmns+\seq}\xspace}
\newcommand{\htmseq}{\texttt{\htmns+\seq}\xspace}
\newcommand{\tsxundo}{\texttt{\htmns+undo}\xspace}
\newcommand{\tsxredo}{\texttt{\htmns+redo}\xspace}
\newcommand{\ccstm}{\texttt{ccSTM}\xspace}
\newcommand{\stm}{\texttt{STM}\xspace}
\newcommand{\sysseq}{\texttt{\sys-seq}\xspace}
\newcommand{\sysredo}{\texttt{\sys-redo}\xspace}
\newcommand{\sysundo}{\texttt{\sys-undo}\xspace}

\newcommand{\htmredo}{\texttt{\htmns+\redo}\xspace}

\newcommand{\undospinlock}{\texttt{undo}\xspace}
\newcommand{\redospinlock}{\texttt{redo}\xspace}
\newcommand{\tsxspinlock}{\texttt{\htmns+spinlock}\xspace}
\newcommand{\tsxundospinlock}{\texttt{\htmns+undo}\xspace}
\newcommand{\tsxredospinlock}{\texttt{\htmns+redo}\xspace}
\newcommand{\sysundospinlock}{\texttt{\htmns+undo}\xspace}
\newcommand{\sysredospinlock}{\texttt{\htmns+redo}\xspace}

\begin{document}
\title{Persistence and Synchronization: Friends or Foes?} 
%\title{Characterizing non-volatile memory transactional systems}
\author{Pradeep Fernando$^1$ \qquad Irina Calciu$^2$ \qquad Jayneel Gandhi$^2$ \qquad Aasheesh
  Kolli$^3$ \qquad Ada Gavrilovska$^1$\\$^1$Georgia Tech \qquad\qquad $^2$VMware Research \qquad\qquad $^3$Penn State\\
\normalsize{\emph{pradeepfn@gmail.com \quad icalciu@vmware.com \quad gandhij@vmware.com \quad aasheesh.kolli@gmail.com \quad ada@cc.gatech.edu}}\vspace{-0.15cm}}
\date{}
\maketitle
\thispagestyle{plain}
\pagestyle{plain}

% !TEX root = paper.tex

\begin{abstract}

Emerging non-volatile memory (NVM) technologies promise memory speed
byte-addressable persistent storage with a load/store
interface. However, programming applications to directly manipulate
NVM data is complex and error-prone. Applications generally employ
libraries that hide the low-level details of the hardware and provide
a transactional programming model to achieve
crash-consistency. 
Furthermore, applications continue to expect
correctness during concurrent executions, achieved through the use of
synchronization.
To achieve this, applications seek well-known ACID guarantees.
However, realizing this presents
designers of transactional systems with a range of choices in how to
combine several low-level techniques, given target hardware features and
workload characteristics.
%The combined property, sought after by multi-threaded applications
%using NVM as main memory, is one that provides both crash-consistency
%and synchronization -- crash-sync-safety. 

In this paper, 
we provide a comprehensive evaluation of the impact of combining
existing crash-consistency and synchronization methods for achieving
performant and correct NVM transactional systems.
We consider different
hardware characteristics, in terms of support for hardware
transactional memory (HTM) and the boundaries of the persistence
domain (transient or persistent caches).
By characterizing persistent transactional systems in terms
of their properties, we make it possible to better understand
the tradeoffs of different implementations and to arrive at better
design choices for providing ACID guarantees. 
%we define crash-sync-safety and explore how it can be
%achieved using a single transactional primitive.
%We evaluate various implementations of such a transactional library
%that provides not only crash-consistency, but also ensures correct
%synchronization of shared data simultaneously. We implement and
%optimize this library for various NVM characteristics and persistence
%domains, and evaluate the cost of crash-consistency and
%crash-sync-safety for a multitude of diverse workloads.
We use both real
hardware with Intel Optane DC persistent memory and simulation to evaluate a persistent version of
hardware transactional memory, a persistent version of software
transactional memory, and undo/redo logging. Through our empirical
study, we show two major factors that impact the cost of supporting persistence in transactional systems:
%crash-consistency and crash-sync-safety: 
the persistence domain
(transient or persistent caches) and application characteristics, such
as transaction size and parallelism.
%We conclude from our experiments
%that combining synchronization with crash-consistency is beneficial to
%applications. 

\end{abstract}

% !TEX root = paper.tex

\section{Introduction}

Emerging Non-Volatile Memory (NVM) technologies, like Intel's 3D XPoint~\cite{3dxpoint},
offer byte-addressability and orders of magnitude faster access to storage than traditional storage technologies.
Their key appeal is that they allow applications to access storage directly using processor load and store instructions rather than relying on a software intermediary like the file system or a database~\cite{pelleychen14}.
However, ensuring that data stored in NVM is always in a safe and recoverable state is both hard and incurs performance overheads~\cite{bpfs,mnemosyne,nvheaps,pelleychen14}.

To ensure data recoverability, application developers have to carefully orchestrate data movement from the volatile to the persistent components in the memory hierarchy, subject to application-specific constraints.
This task is especially complex due to two factors: 
(1)
%varying guarantees on when data is considered persistent and 
NVM applications have very diverse crash-consistency requirements~\cite{whisper}; and
(2)
% diversity in the crash consistency requirements of various NVM applications.
 the \emph{persistence domain} is different across platforms. 
For example, Intel and Micron guarantee that data becomes persistent only when it reaches the memory controller of the NVM device, i.e., the persistence domain of the system includes the memory controller and the NVM devices~\cite{intel2016pcommit}.
We refer to such systems as having \emph{transient caches}.
However, HPE's NVM~\cite{hpnvdimm} guarantees that the entire cache hierarchy is persistent, i.e., the persistence domain includes the entire memory hierarchy.
We refer to such systems as having \emph{persistent caches}.
%Multi-threaded applications require developers to ensure correct synchronization on top of crash consistency.
%These factors make NVM programming hard.

In this context, researchers have proposed various transactional systems that provide the well known ``ACID'' guarantees for NVM applications~\cite{mnemosyne,nvheaps,phtm,phytm,Wang:cal:2015,Joshi:isca:2018,Doshi:hpca:2016}.
These transactional systems significantly simplify NVM application development and leave the complexities of achieving data recoverability on various platforms to the low-level systems software developers.
While these systems all provide ACID guarantees, they go about providing these guarantees in different ways: UNDO vs. REDO logging, software vs. hardware transactions. 
%Different implementations are suited for different applications.
%So, 
Low-level developers designing ACID transaction systems face a bewildering array of choices, with varied performance characteristics that change with the applications and the platform used. 
%For a platform with given hardware features and recoverability requirements, 
For these developers, we aim to answer the question: \textbf{how to quickly explore the design space and arrive at a correct and high-performance implementation of a NVM transactional system?}

Reasoning about implementation details rather than the overall guarantees provided to the user (ACID) helps transaction system developers traverse the design-space more efficiently.
To provide ACID guarantees, the underlying transaction system has to correctly ensure three properties: (1) \emph{crash consistency} - individual transactions are failure-atomic, i.e., after a crash, either all or none of the transaction has persisted, (2) \emph{synchronization} - transactions are correctly isolated from other transactions executed on different threads, and (3) \emph{composability} - the crash consistency and synchronization techniques used compose to provide the overall ACID guarantees, by ensuring that dependent transactions are correctly ordered.
%We coin the term \csty to characterize transaction systems that provide all three of the above properties.
%: crash consistency, synchronization, and their composability.

This new characterization of transaction systems provides a basis to compare different implementations and to identify the right set of \cc and \sync mechanisms for particular applications and hardware platforms.
We perform a detailed characterization study of systems with different implementations (hardware transactional memory (\htm) \cite{Joshi:isca:2018}, 
software transactional memory (\stm) \cite{mnemosyne}, and \undo/\redo logging with locks \cite{pmdk,mnemosyne}) under various persistence domains (transient vs. persistent caches).
We perform our study on real hardware using the recently released Intel's DC Optane Persistent Memory~\cite{dcoptane} and using simulation.
%Finally, we provide an analysis of the relationship of \csty and the best implementation of a transactional programming model for a given NVM system and application.

Our empirical study results in several interesting insights for NVM transaction system developers:
\begin{enumerate}[leftmargin=*]
	\item For all applications, the persistence domain plays the most important role. The overhead of making transactions persistent is considerably lower when caches are persistent.
	\item In systems with transient caches, HTM is the best choice, despite its synchronization costs and required architectural changes. This is due to the high overheads caused by flush and fence instructions required by undo/redo logs, which are elided by HTM. The choice between undo and redo logs depends on the application characteristics and the size of the read and write sets of the transactions.
	\item In systems with persistent caches, the HTM does not require any architectural changes, but its benefit for supporting persistent transactions is reduced, as software logging mechanisms do not require expensive flush and fence instructions anymore. Here, undo logs are the best choice because redo logs suffer from read-indirection overheads.
	\item The overheads of crash-consistency for an HTM are subsumed by synchronization overheads. As applications scale, performance increases despite crash-consistency overheads. When the crash-consistent HTM does not achieve scalability due to aborts, crash-consistent STM ensures this property.
\end{enumerate}

\begin{table*}[t]
    \small
    \centering
    \begin{tabular} [t!]{|l|l|l|l|}
        \hline
         & \textbf{Baseline Tx} & \textbf{Undo Tx} & \textbf{Redo Tx}\\
        \hline
        1 & tx\_begin(); & tx\_begin(); & tx\_begin();\\
        2 & \enskip pA = x; & \enskip log[\&pA] = pA; clwb(log[\&pA]); sfence; pA = x; & \enskip log[pA] = x;\\
		3 & \enskip y = pA; & \enskip y = pA; & \enskip y = (log[pA] $||$ pA); \\
        4 & \enskip pB = z; & \enskip log[\&pB] = pB; clwb(log[\&pB]); sfence; pB = z; & \enskip log[pB] = z;\\
        5 & tx\_end(); & \enskip persist\_write-set(); commit\_log(); & \enskip clwb(log[pA]); clwb(log[pB]); sfence; commit\_log();\\
        6 & & tx\_end(); & \enskip replay\_log(); persist\_write-set();\\
        7 & & & tx\_end();\\
        \hline
    \end{tabular}
    \vspace{0.1cm}
    \caption{\small Undo vs Redo logging; Undo logging suffers from frequent cacheline flushes and sfences while redo logging suffers from read-indirection overheads.}
      \label{tab:undoredo}
	  \vspace{-0.5cm}
\end{table*}

Overall, this paper makes the following contributions:
\begin{itemize}
	\item We characterize persistent transactions to quickly and methodically compare different implementations of NVM transactional systems that provide ACID guarantees.
	\item Using this new characterization, we study the performance of various transaction system implementations on different hardware platforms and for different applications.
	\item We show that there is no one best way to provide ACID guarantees for NVM applications; the best way changes with hardware platforms and application characteristics.
	\item Finally, we believe we are the first work to evaluate these different transactional systems on real 3D XPoint devices.

\end{itemize}

\section{Background}

%\subsection{Non-volatile memory (NVM)}
In order to illustrate the complexity of the design space, 
%We survey the design space by briefly summarizing 
we briefly survey different implementations for crash-consistency
(\S\ref{sec:failureatomic-tx}), for transaction synchronization  (\S\ref{sec:tm}),
and the impact of the hardware persistence domain on the relationship between
the two (\S\ref{ptcache}). 

%%%%%%%%%%%%%%%%%%%%
\subsection{Crash-consistent transactions}
\label{sec:failureatomic-tx}

Crash-consistent (failure-atomic) transactions ensure that a group of updates to NVM locations performed by an application persist atomically, i.e., either all of them are observable or none of them are observable after a failure.
Transactions are specified using \texttt{tx\_begin()} and \texttt{tx\_end()} calls. All the updates to NVM between those two successive calls are guaranteed to persist atomically.
For example, in Table~\ref{tab:nvmtsxcode}, the updates to pA and pB are crash-consistent.
%We discuss possible implementations of \emph{crash-consistency} in Section~\ref{sec:design}.
\emph{Crash-consistency} is generally achieved using undo or redo logging. 

%There are many different ways to ensure the failure-atomicity of transactions and we discuss them in detail in Section XX.

%\subsubsection{\undo logging}
%\label{sec:undo}
\noindent\textbf{undo logging}
 is a crash consistency technique that provides failure atomicity by undo-ing (or rolling back) changes from an aborted failure-atomic transaction.
To be able to roll back changes, \undo logging systems create an \undo log entry \emph{prior} to every update performed within the transaction.
The \undo log entry contains the current value of the memory location/variable that is being updated.
Once the log entry had been created and persisted, only then is the actual memory location/variable updated.
If a transaction succeeds, all the memory locations modified within the transaction are persisted and then a commit message is atomically persisted to the log to invalidate the log entries belonging to the transaction.
If a transaction fails, during the recovery process, all valid log
entries are used to roll back partial changes from a transaction.
Table~\ref{tab:undoredo} illustrates the operations which need to
performed when using different logging techniques. While certain
optimizations may be applicable under some scenarios, the code snippets
represent the steps necessary in the general case. 
As shown in %Table~\ref{tab:undoredo},
the table, \undo logging systems must ensure that: (1) within a transaction, log entries must be created and persisted prior to every update and (2) at the end of the transaction, all memory locations modified within the transaction must be persisted before transaction commit.
 
\begin{table}[b]
	\footnotesize
	\centering
	\begin{tabular} [tbh]{l l l}
		\underline{THREAD-1} & & \underline{THREAD-2} \\
		tx\_begin(); &  & \\
		\enskip \textbf{pA = x;} &  & tx\_begin(); \\
	\enskip pB = y;  & & \enskip \textbf{if (pA == x)}\\
	   tx\_end(); & &  \enskip \quad \textbf{pD = z;} \\
	    &  &\enskip pC = w;\\
		   & &tx\_end(); \\
	  \end{tabular}
	  \vspace{0.1cm}
	  \caption{Threads executing dependent transactions. Correct implementations ensure that 
		  pA persists before pD, crash consistency might be violated otherwise.}
	\label{tab:nvmtsxcode}
\end{table}

%\subsubsection{\redo logging}
%\label{sec:redo}
\noindent\textbf{redo logging} 
is a crash-consistency technique that provides failure atomicity by redo-ing (or rolling forward) changes from committed failure-atomic transactions.
To be able to roll forward changes, \redo logging systems create a \redo log entry for every update within the transaction.
The \redo log entries contain the latest updates while the actual data is maintained at a prior crash-consistent state.
All the read requests for the memory locations updated within the transaction are serviced from the \redo log.
If a transaction succeeds, a commit log entry is created and persisted in the \redo log, marking the commit of the transaction.
In the event of a failure, the \redo log entries of committed transactions are used to roll forward the application's data to its most recent crash consistent state.
The log entries of uncommitted transactions are simply discarded.
Periodically, the redo log can be truncated to reduce read indirections and to reduce the number of redo log entries that have to be applied during recovery.
As shown in Table~\ref{tab:undoredo}, \redo logging systems must ensure that: (1) within a transaction, a \redo log entry must be created for every update within the transaction and read requests to these locations must be re-directed to the log, and  (2) at the end of the transaction, all the redo log entries and a commit log entry must be persisted.

\subsection{Transactional memory}
\label{sec:tm}

Transactional memory~\cite{tm, stm} is used to synchronize the access of multiple threads to 
shared program data. 
%As with failure-atomic transactions,  
Programmers enclose the critical code blocks with \texttt{tx\_begin()} and 
\texttt{tx\_end()} calls.  Transactional memory 
guarantees the atomic execution of a transaction, using speculation. If the runtime detects a conflict with 
another transaction, it aborts one of the transactions, discards its speculative 
state and rolls back its execution to 
the \texttt{tx\_begin()} call. 
%The other transaction is allowed to proceed. 
%During the execution, all updates performed inside the transaction are marked as transactional, and are not 
%visible to other threads. 
%A transaction that successfully commits makes its updates visible to other threads atomically. 
%
Software transactional memory (STM)~\cite{shavitstm} is implemented  
using fine grained locking and write set logging in software. 
%Generally, STMs add 
%two types of overhead (1) write set logging and read set validation in software and (2) conflict detection 
%with other transactions.  
%
Hardware transactional memory (HTM)~\cite{tm} is implemented 
using the
L1 cache to buffer speculative writes and the cache-coherency protocol to
detect conflicts with other threads. 
% Modern processors such as Intel Xeon, have 
%HTM support integrated into their micro-architecture. However, the 
Current
HTM implementations, such as Intel Transactional Synchronization Extensions (TSX),  are best effort -- transactions could 
abort for any reason, such as exceeding the L1 cache capacity, using unsupported instructions, or due to interrupts. 
Therefore, HTMs require a fallback mechanism to ensure progress, usually implemented using locking. 
%Locking is generally used on the fallback path and transactions are 
%required to read the lock when they start the execution to guarantee correct synchronization between the hardware transaction and the 
%fallback path. 

\subsection{Persistent and transient caches}
\label{ptcache}

There is much diversity among the types of NVM technologies that are available on the 
market, as different vendors provide different performance characteristics and persistence guarantees.
For example, HPE offers a battery-backed DRAM solution~\cite{viyojit}. %'s DRAM+SSD NVM solution~\cite{viyojit} comprises a complex system using 
%commodity DRAM and SSDs, together with a battery or uninterrupted power supply (UPS), which ensures 
%that the  data stored in DRAM will be safely transferred to the SSD in case of a power-failure. 
As this design is based on DRAM, the exposed NVM's latency and bandwidth are as good as for 
DRAM. In addition, the battery can extend the persistency domain to the entire memory hierarchy, 
including CPU caches. \emph{\bf Persistent caches} ensure that all modified cache lines are effectively persistent.
%that have not been 
%propagated to NVM will be automatically flushed in case of a power failure. 
On the other hand, Intel and Micron's proposed 3D XPoint technology~\cite{3dxpoint} has higher latency 
and lower bandwidth than DRAM, while the persistent domain includes only the memory 
controller, but not the CPU caches~\cite{intel2016pcommit}. 
%\ic{Do we have a citation for latency and BW numbers?}
In case of a power failure, 
\emph{\bf transient caches} will lose modified data not already written back to the memory controller.
So, the transaction system developer must use the appropriate instruction sequences to ensure that data becomes persistent on different hardware platforms.

\section{Crash-sync-safety}
%\subsection{Crash-sync-safety}
\label{pvs}

In this work, we focus on applications that use a transactional programming model to get ACID guarantees. 
For example, in Table~\ref{tab:nvmtsxcode}, updates within each transaction need to provide
all or nothing semantics when the data gets to NVM. 
Providing ACID guarantees requires that the transactional system correctly implement three components: (1) \emph{crash-consistency} (also called failure-atomicity), which ensures 
all-or-nothing behavior of uncommitted transactions when a failure happens and the validity of the data after the failure (atomicity and consistency) (2) \emph{synchronization}, which ensures 
that partial updates are not observable by other concurrently running transactions (isolation), and (3) \emph{persistence} of the committed transactions in the correct order, which ensures 
that committed transaction 
are made durable and that the correct dependencies between transactions are maintained (durability).
Note that crash-consistency is a property of uncommitted transactions, which guarantees that on a failure, a transaction will either abort, leaving no side-effects, or 
will commit, finishing its entire execution. In contrast, persistence is a property of committed transactions, guaranteeing their permanence in case of a crash, as well as 
that dependent transactions' effects are all visible in the correct order. 
We call a correct implementation of the above three properties that ensures ACID guarantees \cs.

Programmers identify regions of code within their applications as transactions using \texttt{tx\_begin()} and \texttt{tx\_end()}, and are assured of the failure-atomicity of the updates within any transaction.
Furthermore, all the updates within the transaction become atomically visible to any other thread in the system 
once persisted, 
and conflicting transactions (transactions accessing  common memory locations with at least one of the accesses being a write) execute in isolation.
So, each transaction behaves as both a traditional transactional memory transaction and also a failure-atomic transaction.

Developers have a wide variety of choices for \cs transactions, and choosing between these different options 
depends on a variety of factors, such as the persistence domain, and the application characteristics. 
To further complicate matters, some mechanisms offer some of the guarantees, but not all, and developers need to 
carefully mix and match techniques to ensure correctness. 
For example, \undo and \redo logging can be used to implement crash-consistent transactions for single-thread applications, but do not ensure 
the correct synchronization of multi-threaded applications, forgoing isolation. 
Conversely, locking can be used to provide correct synchronization for 
multi-threaded applications, but cannot ensure persistence for these transactions in case of a failure, forgoing durability, nor 
crash-consistency, forgoing atomicity and consistency. 
Transactional memory provides correct synchronization for 
multi-threaded applications, as well as atomicity and consistency, but cannot ensure persistence for these transactions in case of a failure, forgoing durability.

\section{Crash-sync-safe Transactions}
\label{sec:design}

This section describes different implementations of a transactional library that ensures \csty in detail.
First, to achieve proper synchronization, transactions may be implemented using one of three broad approaches: (1) Hardware Transactional Memory (HTM), (2) Software Transactional Memory (STM), or (3) global locking.
Each of these approaches can further be extended to additionally provide crash-sync-safety for transactions.
Depending on whether the system has transient or persistent caches, the implementation details will vary.
Next, we describe these different implementations (Table~\ref{table:tl-impl}).

\begin{table}[t]
\centering
\resizebox{\columnwidth}{!}{%
\begin{tabular}{|l|c|c|c|c|c|}
                \hline
                & \multicolumn{2}{c|}{ST -- CC} & {MT -- Sync.}  & \multicolumn{2}{c|}{MT -- CSS} \\ \cline{2-6}
                & {TC} & {PC} & & {TC} & {PC} \\ \hline
                {seq} & {\ding{55}} & {\ding{55}} & {\ding{55}} & {\ding{55}} & {\ding{55}} \\ \hline
                {HTM+seq} & \multirow{2}{*}{\ding{55}} & \multirow{2}{*}{\ding{55}} & \multirow{2}{*}{\ding{51}} & \multirow{2}{*}{\ding{55}} & \multirow{2}{*}{\ding{55}} \\
                {(+spinlock)} &&&&&\\ \hline
                {undo/redo} & \multirow{2}{*}{\ding{51}} & \multirow{2}{*}{\ding{51}} & \multirow{2}{*}{\ding{51}} & \multirow{2}{*}{\ding{51}} & \multirow{2}{*}{\ding{51}} \\
                {(+spinlock)} &&&&&\\ \hline
                {HTM+undo/redo} & \multirow{2}{*}{approx.} & \multirow{2}{*}{\ding{51}} & \multirow{2}{*}{\ding{51}} & \multirow{2}{*}{approx.} & \multirow{2}{*}{\ding{51}} \\
                {(+spinlock)} &&&&&\\ \hline
                {ccHTM+undo/redo} & \multirow{2}{*}{\ding{51}} & \multirow{2}{*}{N/A} & \multirow{2}{*}{\ding{51}} & \multirow{2}{*}{\ding{51}} & \multirow{2}{*}{N/A} \\
                {(+spinlock)} &&&&&\\ \hline
                {STM} & {\ding{55}} & {\ding{51}} & {\ding{51}} & {\ding{55}} & {\ding{51}} \\ \hline
                {ccSTM} & {\ding{51}} & {N/A} & {\ding{51}} & {\ding{51}} & {N/A} \\ \hline
\end{tabular}
}
\vspace{0.1cm}
\caption{Crash-consistency and crash-sync-safety implementations for single- and multi-threaded applications. ST: Single-threaded, MT: Multi-threaded, CC: Crash-consistent,
        Sync: Synchronization, CSS: Crash-sync-safety, TC: Transient caches and PT: Persistent caches. Techniques evaluated for single-threaded applications need to provide only 
        crash consistency. Techniques evaluated for multi-threaded applications provide synchronization too, by using a spinlock where necessary. We note that the HTM+undo/redo 
        implementations for transient caches are only approximating a crash-sync-safe solution.}
\label{table:tl-impl}
\end{table}

\subsection{Crash-sync-safe HTM}
\label{sec:cchtm}

Hardware Transactional Memory (HTM)  offers atomicity and isolated transactions for volatile memory.
With persistent memory systems, HTM implementations can be extended to ensure that they become \cs.
Designing crash consistent HTM (ccHTM) requires augmenting HTM with a separate \undo/\redo log in persistent memory~\cite{phtm} and logging data modifications within a transaction.
It is important to note that the software fallback path must also be made crash consistent through appropriate logging.
In the event of a failure, the ccHTM logs can be used to restore the application's persistent data to the most recent consistent state.
While many different ccHTM implementations have been proposed recently~\cite{phtm,phytm,Wang:cal:2015,Joshi:isca:2018,Doshi:hpca:2016}, to the first order, they are all similar.
In this work, we developed and implemented our own ccHTM design %(\S\ref{sec:ntsx}), 
as a representation of the prior proposals. 

\noindent{\bf Transient caches.} For transient caches
%Our 
our \sys implementation augments a regular HTM~\cite{inteltsx} % \ada{specific reference that's closest? } -- added reference for intel RTM
with support for failure-atomicity when modifying data in NVM.
Apart from the usual read/write set tracking employed in HTM designs, ccHTMs employ a separate write-set log in NVM to maintain failure-atomicity.
To update the NVM log, we extend the hardware to issue write-combined, non-temporal stores (those that bypass the cache hierarchy, like x86's \texttt{movnt}~\cite{intel-flush}) for every write within a transaction. These writes are non-transactional operations~\cite{logtm}, so they do not become part of the transaction's write set.
Note that reads and writes that are part of the transaction use the regular temporal load/store instructions and are served from the CPU caches.
We use \redo logging in our \sys implementation, but \undo logging would be similar. When used inside a hardware transaction, 
the \redo log does not suffer from read indirection, because the values can be found as speculative values in the L1 cache. 
At commit time, we first ensure that the log writes are persistent, then atomically persist a log commit message in the NVM log, then make the transaction's updates visible to other cores in the system.
Overall, transactions first attempt an execution as a failure-atomic hardware transaction.
However, if a hardware transaction aborts, the fallback path involves acquiring a global lock and executing pessimistically using a software undo/redo log.
Next, we describe in detail the various aspects of our \sys implementation.

\begin{figure}[htb] 
	\centering
	\includegraphics[keepaspectratio,width=0.8\linewidth]{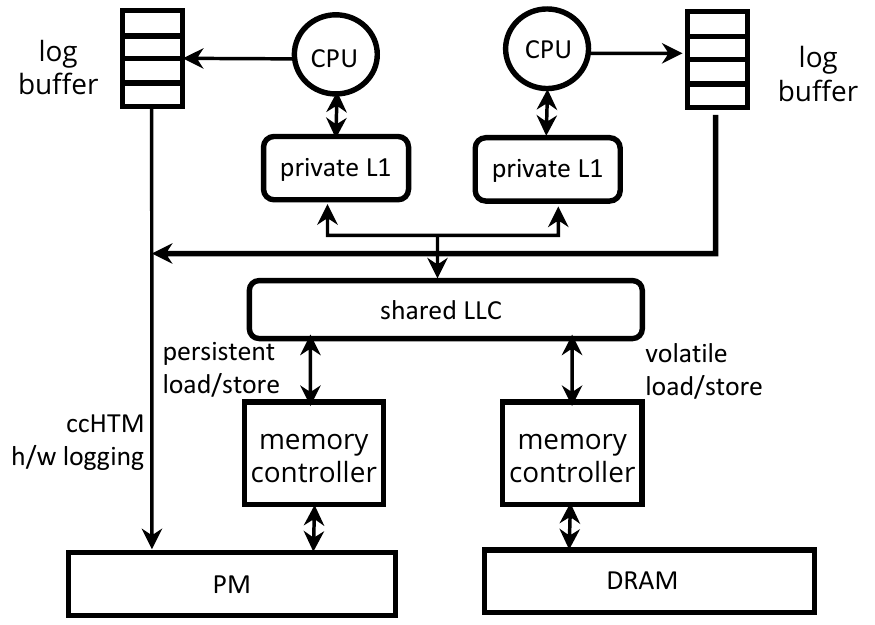}
	\caption{\small Design of \sys} 
	%\vspace{-2mm}
\end{figure}

\noindent{\bf Persistent write-set logging.} HTM runtimes keep track of the read/write sets of
the executing transaction. 
Writes executing inside a transaction are held in the L1 cache in speculative state, isolated
from the rest of the memory hierarchy until the successful completion and commit of the transaction. 
%\ada{there is an earlier claim that there are improvements to logging techniques. clarify where does prior system (with reference) end and improvements begin. is it just related to the non-temporal stores mentioned above?} -- claim removed
Similar to HTMs, \sys issues writes in the transaction to the L1 cache.
In addition, \sys intercepts each write and augments it with a hardware based non-temporal log-write request into a
thread-local NVM log.
So, every write within a transaction results in a temporal write to the L1 cache and a non-temporal log write.
The NVM log write is asynchronous to the intercepted transactional write, thus has minimal
impact on the transaction's critical path. 
The 
\sys log is durable and atomic updates do not suffer from inherent read indirection overheads of logging, as incoming read
requests are being served directly from the L1 cache.

\noindent{\bf Transaction commit.} An \sys transaction comprises both volatile state 
(the cache lines held speculatively in the L1 cache) and persistent state (the \sys-log).
Thus, the \sys transaction commit sequence differs from a traditional HTM commit.
The \sys commit sequence includes: (1) updating the book-keeping structures of speculative cache-lines, (2)
failure-atomic write-set log commit on NVM, and (3) atomically releasing the speculative
cache-lines to the rest of the memory hierarchy.
It is important to note that \sys commit operation combines two commit phases --
a persistent memory commit, in the form of \sys-log commit, and
a volatile memory commit, in the form of speculative cache-line unlocking.
A \sys log commit involves two \texttt{sfence} instructions. 
The initial \texttt{sfence} drains the buffered asynchronous log writes to the \sys-log.
Next we atomically persist/update the \sys-log's tail-index with the latest log-entry index value, followed by
another \texttt{sfence}. 
The \sys-log's tail-index update doubles as a commit-flag entry and enables fast log truncation.
%Next we issue the commit flag append to the \sys-log followed by another \texttt{sfence}.
Once the transaction has been committed in NVM, the volatile commit is performed by atomically moving the affected cache-lines out of the speculative state.
Once the volatile commit is performed, the transaction has successfully completed.

\noindent{\bf Transaction abort.}
Similar to HTM aborts, \sys aborts may be triggered due to (but not limited to) a 
 load/store on another thread that conflicts with the current transactions' write/read
set, OS interactions like system calls or context switches, L1 cache capacity overflow.
In addition to all of the HTM abort causes, \sys transactions abort if the runtime
runs out of \sys-log space during hardware logging. We introduce a new abort flag
called \texttt{NO\_LOG\_SPACE} to capture this abort cause.
A transaction abort includes (1) discarding speculative L1 cache-lines, 
and (2) invalidating \sys-log appends. 
We rely on existing HTM capabilities to achieve (1). We do not explicitly invalidate 
\sys-log entries as they remain invalid till \sys-log's tail-index update.

%\subsection{Fallback path} 
\noindent{\bf Fallback path.}
Similarly to hardware transactions, \sys transactions are best effort -- the transactions are not guaranteed to complete.
\sys transactions use regular write-ahead-logging (WAL) on the fallback path to ensure persistence. 
The fallback path transactions use either \undo or \redo logging while HTM transactions use \redo logging.
In addition, a global lock ensures the synchronization between 
the fallback path software transactions and \sys hardware transactions. To ensure isolation, the hardware 
transactions read the global lock as soon as they start executing, which makes them abort if another 
thread acquires the lock.

%\subsection{Log truncation}
\noindent{\bf Log truncation.} %\ada{Only during fallback?} -- the text describes ccHTM log truncation. Fallback is same.
\label{log-truncate}
We truncate the transaction logs (both ccHTM and fallback path) at the end of each transaction -- eager log truncation.
With \redo logging, log truncation involves first persisting the cache-lines modified as part of the transaction and then invalidating the transaction's log entries.
We truncate the ccHTM transaction logs as part of the transaction commit step, i.e., once the NVM log of the transaction is committed, 
we perform the following steps: (1) issue \clwb requests to all the cache-lines in the write-set, (2) issue an \sfence to ensure their writeback, 
(3) issue a non-temporal update request to atomically reset the \sys-log's tail-index to truncate/invalidate all the previously written log entries, and
%(3) issue a non-temporal ``truncate'' marker to the NVM log  to mark the truncation of the log entries that belong to the transaction, and 
(4) issue another \sfence to ensure that the update request has been persisted.
Once these four steps are performed, the volatile commit of the transaction is carried out.
We also truncate the logs for the transactions executed in the fallback (software) path as soon as they commit.
It is important to note that since both the fast path and slow fallback path employ log truncation, 
it is feasible to employ different logging techniques in the different paths.
For example, it is possible to use \redo logging in the fast path and \undo logging on the fallback path.
This design approach allows us to evaluate \cc mechanisms that use different logging techniques on the different paths.
Furthermore, this eager log truncation approach, relives our \sys implementation of the burden of tracking the execution order of different transactions in their respective NVM logs, as is necessary in other prior approaches~\cite{Wang:cal:2015}.

\noindent{\bf Persistent caches.}
In systems with persistent caches, speculatively updated ccHTM cachelines are persistent as soon as they are atomically released. (when they made visible in L1 cache).
However, transactions executing in fallback path still need atomic updates in the form of undo/redo logging.
So, regular HTM implementations can be augmented with a fallback path log and can ensure crash-sync-safety with no additional changes to the HTM.

\subsection{Crash-sync-safe STM}
\label{sec:ccstm}

Software Transactional Memory (STM) offers atomicity and isolated transactions for volatile memory.
All the data modifications made within the transaction are made visible to other threads atomically when the transaction commits.
If the transaction aborts, none of the data modifications made within the transaction become visible.
STM implementations track the read and write sets of individual transactions to ensure transaction atomicity.
Further more, they provide transaction isolation by detecting conflicting transactions that modify at least one common memory location and aborting some of them as necessary.

With persistent memory systems, STM implementations can be extended to ensure that they become crash-sync-safe, i.e., data modifications within a transaction will persist atomically and 
the dependencies are handled properly (\S\ref{pvs}).
There are two broad approaches to designing crash consistent STM (ccSTM): (1) augment STM with a separate \undo/\redo log in persistent memory~\cite{mnemosyne,nvheaps} or (2) repurpose the write sets already maintained as part of the STM implementation to also function as a \undo/\redo log.
In the event of a failure, the ccSTM logs can be used to restore the application's persistent data to the most recent consistent state.
In this work, we concentrate on ccSTM designs that maintain a separate \undo/\redo log, similarly to \cite{mnemosyne}.

\noindent{\bf Transient caches.}
In systems with transient caches, in order to make sure their log entries are persistent (\undo or \redo), ccSTM designs have to write back the log entries from the processor caches to the memory controller.
Furthermore, log entries have to be written back as per the ordering constraints of the logging mechanism employed (as discussed in \S~\ref{sec:failureatomic-tx}) using carefully orchestrated \clwb, \sfence, and non-temporal store instructions (e.g., \texttt{movnt}).

\noindent{\bf Persistent caches.}
However, in systems with persistent caches, ccSTM log entries are persistent as soon as they are created (when they reach the L1 cache).
So, regular STM implementations also ensure crash consistency with no additional changes in systems with persistent caches.

\subsection{Crash-sync-safe locking}
This implementation of transactions acquires a global spinlock at the beginning of every transaction and releases it at the end of every transaction.
While this naive implementation suffers from frequent false conflicts for multi-threaded applications, it does offer one advantage.
It is a very light-weight approach when no concurrent transactions are executed by an application, an extreme case of which is a single-threaded application.
Mostly, we use this design point for the sake of completeness in our \csty design space analysis.
While global locking achieves proper synchronization, to achieve \csty, transactions are usually extended with either \undo or \redo logging, which we describe next.

\noindent{\bf Transient caches.} %\ada{repeated verbatim from 4.1.1, in Persist. caches too.} 
In systems with transient caches, data can be considered persisted only once it has been written back from the volatile cache hierarchy to the memory controller using one of \clflush, \clflushopt, \clwb instructions.
Further more, some of these instructions are non-blocking, so applications need to issue a subsequent \sfence to ensure that the instructions have been fully executed and the associated data is actually persistent.

\undo logging systems have to ensure that log entries are persistent before they can allow actual memory locations to be modified within a transaction.
As shown in Table~\ref{tab:undoredo}, \undo logging systems use a combination of \clwb and \sfence instructions prior to every data update, i.e., every store instruction.
This frequent use of blocking \sfence instructions could result in severe performance degradation.
\redo logging systems have to ensure that all the redo log entries and the commit log entry are persisted by the end of a transaction.
Further more, the commit log entry may persist only after all the redo log entries have been persisted.
As shown in Table~\ref{tab:undoredo}, \redo logging systems use a combination of \clwb and \sfence instructions within a transaction.

\noindent{\bf Persistent caches.}
However, in systems with persistent caches, data is considered persistent as soon it has been written to the L1 data cache.
So, no writeback of data to the memory controller is necessary on such machines.
On systems with persistent caches, \undo logging implementations need to ensure that log entries are created before the data update, while \redo logging implementations need to ensure that redo log entries are created for every update within the transaction and that the commit log entry is created before the completion of the transaction.
Since x86 systems guarantee TSO, the program order of stores ensures that the stores belonging to the log entry creation and data update are performed in order, without the need for any intervening \sfence or \clwb instructions.
For example, with persistent caches on an x86 machine, all the \clwb and \sfence instructions shown in Table~\ref{tab:undoredo} become obsolete.
However, for systems with a weaker memory model (e.g., ARM), an appropriate \fence instruction is necessary to ensure that the stores are executed in program order.
Persistent caches significantly improve the performance of \undo/\redo logging systems as they eliminate expensive \clwb and \sfence instructions.

%\input{implementation}
% !TEX root = paper.tex

\section{Implementation and evaluation methodology}
\label{sec:methodology}
We want to understand the overheads of crash-consistency and crash-sync-safety, as well as 
what is the best implementation of a transactional library that provides these properties, 
given various NVM characteristics, persistence domains, and workload characteristics.
To do so, we compare the performance of different transactional library implementations along the following axes:
(1) Persistence domains of the system.
Specifically, we consider systems with \emph{persistent}
and \emph{transient} caches (\autoref{ptcache}).
(2) 
%Number of threads in the applications, i.e. 
Single-threaded vs multi-threaded applications.
%As described in  
Single-threaded applications just require crash consistency while multi-threaded applications
require crash-sync-safety (\autoref{pvs}). 
%Therefore, single-thread and multi-threaded applications need to be evaluated separately.
(3) Evaluation platform -- real hardware with Intel Optane DC NVM or an architectural simulator.

\noindent{\textbf{Real Hardware vs Architectural Simulation.}}
To evaluate \cchtm, we use two different platforms: (1) bare-metal hardware with TSX and Intel Optane NVM and 
(2) SESC, a cycle-accurate simulator.
While neither approach allows us to accurately evaluate 
\cchtm, they complement each other and provide
a comprehensive analysis of the competing mechanisms.
For example, simulation accurately models the proposed hardware changes not
possible with TSX.
However, real hardware more accurately factors in transaction abort rates 
introduced due to system jitter (background
activities, thread context switches, cache capacity constraints) and the latency and bandwidth constraints of real NVM DIMMs. 

\noindent{\bf Intel Optane NVM hardware testbed:} 
%\ada{Pradeep, check
%  to make sure this is correct:} -- verified and updated
We use a Intel Xeon server (Cascade Lake microarchitecture) with 96 cores over 2 NUMA sockets.
%cores each, running a single hardware thread per core. 
We use Fedora Linux as the OS.
The Intel processor supports restricted transactional memory (\texttt{rtm}) and  the cache-line-write-back instruction (\texttt{clwb}).
Each processor
socket has access to 375 GB of DRAM and 756 GB of Intel Optane NVM,
configured in Direct Access Mode~\cite{steve:report}. 
%Both memories
%are directly accessible via load/store instructions. 
The
NVM memory is managed by a DAX supporting file-system, hence applications have to explicitly map NVM memory into their process address 
space prior to using the NVM. Therefore, we modify our applications to
explicitly allocate all memory dynamically from the NVM address space
using the \texttt{libvmem} allocator from  the Persistent Memory
Development Kit (PMDK)~\cite{pmdk}. We allocate persistent memory (mmap) from the closest NVM DIMM (NUMA aware).
We bind each of the application threads to a compute core.
Thread binding prioritizes the compute cores within the same NUMA socket and assigns compute cores from a different socket only when an application
uses up all the cores in the current socket.
Out of five runs, we report the mean of the middle three runs.

\begin{table}[tb!]
	\centering
	\footnotesize
	\begin{tabular}[htbp]{| c | c |}
		\hline
		\multirow{2}{*}{Processor} &  16 cores \@ 1 GHz \ \\ & connected via bus-interconnect \\ \hline
		\multirow{2}{*}{L1 cache Ins \& Data} & 64 KB per core / 64 B cacheline/ \\ & 8-way set associative \\ \hline
		\multirow{2}{*}{L2 cache Ins/Data} & 256 KB per core/ 64 B cacheline/ \\ & 8-way set associative/ hit-latency 18 cycles \\ \hline
		\multirow{2}{*}{L3 cache} & 16 MB shared/ 64 B cacheline/ \\ & 16-way/ hit-latency 34 \\ \hline 
		{Coherence protocol} & MESI  across L2 caches \\ \hline
		{NVM log size} & 10 MB per core/thread \\ \hline
%		{DRAM} & hit-latency 250 cycles \\ \hline
		{NVM r/w latency} & 250/750 cycles \\ \hline
	\end{tabular}
	\vspace{0.1cm}
	\caption{Simulator config, adopted from~\cite{pleasetm}}
	\label{table2}
\end{table}

\begin{comment}

\begin{table*}[tb!]
\centering
\footnotesize
\begin{tabular}{l|l|l}
	{\bf Workloads} & {\bf Description} & {\bf Transaction Size}\\
	\hline
	{{Vacation}} & {Online transaction processing system to manage travel reservations} & {{Large:} up to 5000 loads and 5000 stores}\\
	{{Kmeans}} & {Clustering application} & {{Small:} up to 100 loads and 20 stores}\\
	{{Ssca2}} & {Graph processing application from computational biology} & {{Small:} up to 10 loads and 2 stores}\\
	{{Labyrinth}} & {Maze route planning application} & {{Very large:} up to 500K loads and 200 stores}\\
	{{Intruder}} & {Signature-based network intrusion detection system} & {{Moderate:} upto 800 loads and 350 stores}\\
	{{Genome}} & {DNA assembling application} & {{Moderate:} up to 2000 loads and 1200 stores}\\
	{{Ctree}} & {Persistent memory, crit-bit tree. Supports both point and range queries.} & {Small transactions.}\\
	{{Hashmap}} & {Peristent memory, hashtable. } & {Mostly small transactions.} \\

	\hline
\end{tabular}
\caption{Description of workloads}
\label{tab:workloads}
\end{table*}

\end{comment}

\noindent{\bf Simulator:}
We implemented \cchtm as an extension to SESC-HTM~\cite{pleasetm},
which emulates the instruction
behavior(commit, abort, etc.) within the HTM\_begin() and HTM\_end() code regions and passes them to
a back-end timing module for simulation.
We augment the writes happening within a HTM transaction with an asynchronous, non-temporal log-write to the NVM resident log,
in addition to the temporal L1 cache write.
Furthermore we implement the \clwb and \sfence instructions necessary for the
correct functioning of software-based crash-consistency mechanisms.
%\ic{How many lines of code?}
Table~\ref{table2} lists
the configuration of the various hardware structures
modeled in our simulator.
Since SESC was designed 
%primarily 
for MIPS, we cross compile the STAMP
benchmarks and the 
\sys library into a MIPS binary.

\noindent{\bf Workloads.}
\label{sec:workloads}
We use two benchmarks from the PMDK project~\cite{pmdk}, namely C-tree and Hashmap. These two benchmarks
implement a persistent crit-bit tree and a hashmap. We port these two applications to use different persistent memory transactional 
mechanisms. We run the \texttt{pmembench} workload generator provided with PMDK and use workload parameters from ~\cite{whisper}.
In addition, we use the transactional applications from STAMP~\cite{stamp}, a popular
benchmark suite used by others to evaluate libraries for NVM~\cite{whisper, Giles:spaa:2017, Giles:ismm:2017}.
We augment transactions with crash-consistency, on top of
the atomicity, consistency, and isolation guarantees already provided.
To better understand these workloads, we instrumented the simulator to count the load/stores for each transaction. 
%We summarize the results in Table~\ref{tab:workloads}. 

 % !TEX root = paper.tex

 % !TEX root = paper.tex
 
\section{Evaluating crash-sync-safety}

In this section, we seek to understand the cost of implementing 
crash-sync-safety (\S\ref{pvs}) in various ways. 
To do so, we evaluate multi-threaded applications that provide both \cc and \sync. We use the \ccms in Table~\ref{table:tl-impl}. 
We use a spinlock to ensure correct synchronization for the \undo/\redo logs and on the fallback path of the \htm. 
We want to answer the following questions:
(1) What is the most efficient implementation of crash-sync-safe transactions? To answer this question, we compare HTM-based crash-sync-safe transactions with STM-based crash-sync-safe transactions and 
with undo/redo logging using a spinlock. 
(2) What is the overhead of achieving crash-sync-safety? To answer this question, we compare to a sequential implementation baseline, with 
no \cc and no \sync. 
(3) What is the overhead of crash-consistency for multi-threaded applications that are properly synchronized? To answer this question, we compare with a non-crash-consistent 
baseline (HTM+spinlock).
(4) How does the persistence domain of the NVM influence the results? To answer this question, 
we consider two different NVM devices: one with   
transient caches (\S\ref{sec:mt-transient}) and one with persistent caches (\S\ref{sec:pcache}). Current HTM for systems with transient caches ensure 
proper \sync, but not \cc, so our real hardware evaluation on transient caches only approximates a crash-sync-safe implementation based on HTM. Therefore, we use 
simulation to properly evaluate the overheads of the crash-sync-safe HTM. 

\noindent\textbf{Summary of crash-sync-safety results.}
We evaluate multiple transactional implementations on real hardware with the new NVM devices, as well as using an architectural simulator. 
Our results are summarized below. 
(1) We find that ccHTM consistently outperforms other transactional implementations, by 0.06X-30X (at 8 threads) for transient caches, and by 3X on average for persistent caches. 
The only exceptions are applications which are known for being problematic for hardware transactions, i.e., with 
large read or write sets that overflow the cache, or with unsupported instructions that always abort the hardware transactions. Therefore, extending HTM with \cc for transient caches is the most promising solution 
to provide crash-sync-safe transactions. 
The simulation results show that making the HTM crash-consistent does not add significant overhead compared to a non-crash-consistent HTM (HTM+spinlock).
For persistent caches, current HTMs (e.g., TSX) are already crash-sync-safe. 
(2) When using ccHTM to achieve crash-sync-safety, it comes almost for free. The overheads of crash-consistency are subsumed by synchronization overheads and, as applications scale, performance increases compared to single-thread 
execution. When the ccHTM implementation does not achieve scalability due to aborts, the ccSTM still ensures this property. 
(3) If we disregard scalability improvements given by running multiple threads, we can measure the cost of crash-consistency compared to a non-crash-consistent solution that still ensures the synchronization. On average, HTM+undo(redo) is 
2.3X (2.4X) slower than HTM+spinlock (for 4 threads), for transient caches. When caches are persistent, this cost becomes negligible. 
(4) In multi-threaded applications, the persistence domain still plays a very important role, but the results are not as dependent on it as they are for the single-threaded applications. The overhead of crash-consistency is considerably 
lower when caches are persistent. In addition, HTM is \cc out of the box, so no changes are necessary.

%In particular, we see that the notion of crash-sync is indeed useful, as the cost of the \cc is counter-balanced by the improvements in scalability when we combined \cc with ensuring correct \sync using \htm or \stm. The \htm works best, as long as the transactions can fully execute without overflowing the cache, using unsupported instructions or incurring too much contention. In such scenarios, however, the \stm can get very good performance, as it improves the conflict resolution of transactions, ensuring that more transactions can succeed. These results suggest that a crash-consistent hybrid transactional memory could combine the benefits of the \htm and \stm for both transient and persistent caches, resulting in the best way to minimize the cost of crash-sync. 

%\vspace{5cm}
\subsection{Transient CPU caches}
\label{sec:mt-transient}

\begin{figure*}[tp] 
\centering
	\includegraphics[width=\linewidth]{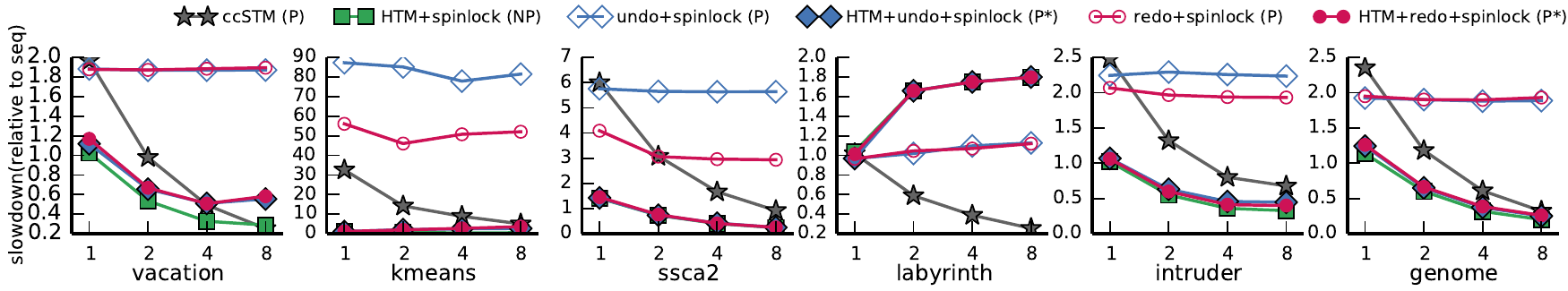}
	\caption{\small TSX-enabled hardware with real NVM and transient caches by number of threads (X axis). 
	(P) crash-consistent; (NP) not crash-consistent; 
	(P*) approximates crash-consistent solution. } \label{fig:Mthr-real} 
\end{figure*}

We use our real test-bed and architectural simulator to evaluate the cost of crash-sync for transient caches.

\noindent{\textbf{Real.}}
Figure~\ref{fig:Mthr-real} shows the scalability of the various approaches outlined above with varying number of threads, on real hardware, using Intel TSX for the \htm.  
\tsxundospinlock (\tsxredospinlock) outperforms \undospinlock (\redospinlock) for vacation, kmeans, ssca2, intruder and genome by $3.6\times$ ($3.7\times$), $30.8\times$ ($18.3\times$), $13.3\times$ ($7.1\times$), $0.6\times$ ($0.6\times$) and $4.9\times$ ($4.8\times$), respectively, at 4 threads. The only exception is labyrinth, where many transactions overflow the cache and abort, ending up executing on the fallback path (\undo/\redo), serialized by a spinlock. 

Compared to the \ccstm, \tsxundospinlock (\tsxredospinlock) is faster for vacation, kmeans, ssca2, intruder and genome by 
$1\times$ ($1\times$), $3.5\times$ ($3.2\times$), $3.9\times$ ($4.0\times$), $1.7\times$ ($1.9\times$) and $1.6\times$ ($1.5\times$) respectively and slower on labyrinth by 53.06\% (53.10\%) at 4 threads. Choosing between software and hardware transactions largely depends on the workload, especially the size of the transactions, conflict rate and usage of TSX-unsupported operations. \ccstm performs better on workloads with larger transactions (e.g., labyrinth) or more contention and scales better to a larger number of threads. We attribute this behavior to its better conflict resolution. However, for small transactions, the software conflict detection and resolution of the \ccstm introduces too much overhead, which is greatly reduced by the simpler hardware-based requester-wins policy of the \htm. 

We approximate the cost of \cc for multi-threaded applications by comparing to \tsxspinlock, which suffers from the overheads of \sync, but not \cc. \tsxundospinlock and \tsxredospinlock are, on average, $2.3\times$ and $2.4\times$ slower than \tsxspinlock for all workloads (at 4 threads).

% undo vs. redo w/ and w/o htm 
As in the single-thread applications, the choice between \undo and \redo logs greatly depends on the workload characteristics. However, when we use \htm on the fast path, the differences between \undo and \redo logs on the fallback path are significantly diminished with \tsxundospinlock and \tsxredospinlock resulting in similar performance. 

% ccstm vs. stm  
We compare the \ccstm with an \stm to  understand the cost of \cc for the \stm. \ccstm is at most $8.2\times$ slower than an \stm with no \cc for 1 thread and at most $7\times$ slower for 8 threads. We see that while \cc definitely adds a noticeable overhead, it does not impact the scalability of the original \stm. Moreover, the difference between \ccstm and \stm decreases with increasing the number of threads, 
%due to a batching effect of the fences employed for \cc. 
showing that the overhead of fences is amortized between multiple concurrent threads.
%\ak{What is the batching effect of the fences?}
%\ic{AK, is this better?}

% the cost of crash-sync - compared to seq
% cost of crash-sync using undo/redo
% cost of crash-sync using htm/stm 
% combining cc and sync results in a much lower crash-sync cost for multi-threaded applications. 
Finally, the cost of \cc does not tell the entire story, as we provide both \cc and \sync for multi-threaded applications. 
Thus, we also measure the cost of \csty by comparing with a baseline with no crash-consistency and no synchronization (\seq). 
While crash-consistency adds overhead compared to volatile in-memory execution, efficient synchronization often improves performance by 
enabling the application to scale to multiple threads. Therefore, the cost of \csty is much lower than the cost of crash-consistency 
when we can pair efficiently \sync and \cc, using \stm or \htm. 
However, when we use distinct methods for the two, the overheads compose and the cost of crash-sync is much higher, as exemplified by \undo/\redo using spinlocks being
$.9\times/1.4\times$\ and $2.7\times/2.9\times$ expensive than \texttt{ccHTM+undo/redo} and \texttt{ccSTM} on-average.
%\ic{Add some numbers here}

As in \S\ref{sec:st-transient}, these numbers are an approximation of \cchtm results, as only the fallback path is crash-consistent. We evaluate a full-fledge \cchtm in the simulator in the next section.

\begin{figure*}[tp] 
\centering
	\includegraphics[width=\linewidth]{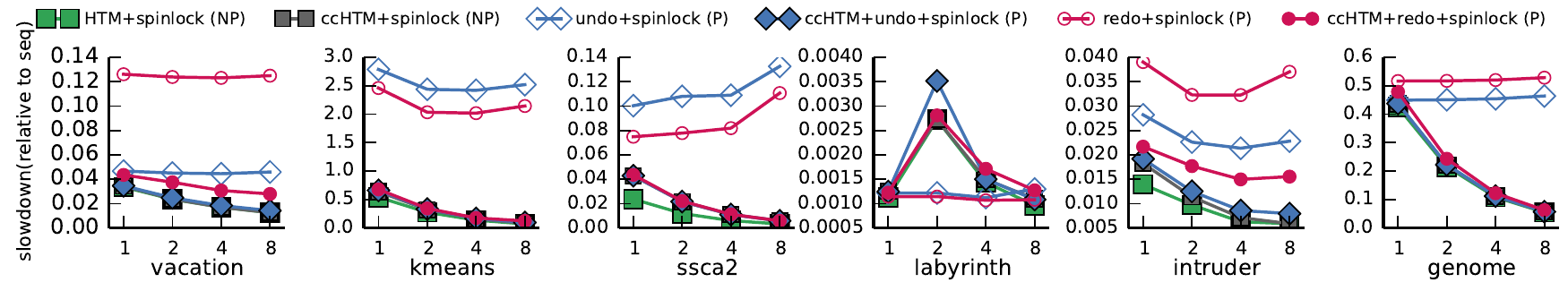}
	\caption{\small Simulation, transient caches by number of threads (X axis). (P) crash-consistent; (NP) not crash-consistent; 
	(P*) approximates crash-consistent solution. }	
	\label{fig:Mthr-sim} 
\end{figure*}

\noindent{\textbf{Simulated.}}
Figure~\ref{fig:Mthr-sim} shows \cchtm results using simulation. We use this to confirm that adding \cc on the \htm fast path does not hinder its performance. 
Once again, the general trends from real hardware largely hold in the simulated environment as well: 
(1) \sysundospinlock/\sysredospinlock comfortably outperform 
\undo/\redo. For example, for Vacation benchmark with 4 threads, the improvements are $3\times$ and $6\times$ respectively.
The only exception to these general trends are seen in the case Labyrinth, where  
 \undospinlock and \redospinlock perform better than HTM based approaches due to the high transaction abort rates inherent to the workload.
 %\ic{This doesn't correspond to the figure anymore.}
(2) \undospinlock/\redospinlock exhibit 
the highest overheads and poor scalability due to the spinlock.
(3) All \ccms increase 
execution time over the non-crash consistent baseline, \tsxspinlock. 
However, on-average \texttt{ccHTM+spinlock} increases execution time by only 8\% compared to \tsxspinlock.
%\ic{compare ccHTM+spinlock with HTM+spinlock}
%
The simulation results differ from the real hardware results mainly in the scalability showed by \cchtm. 
%the HTM based approaches generally scale well with increasing number of threads, 
%(except for a couple of outliers, like Labyrinth for two threads), 
%while that was not true in the case of the real hardware.
This difference comes from the system events that occur in the real systems, but are hard to model in a simulation environment.

	\begin{figure*}[tp] 
	\centering
	\includegraphics[width=\linewidth]{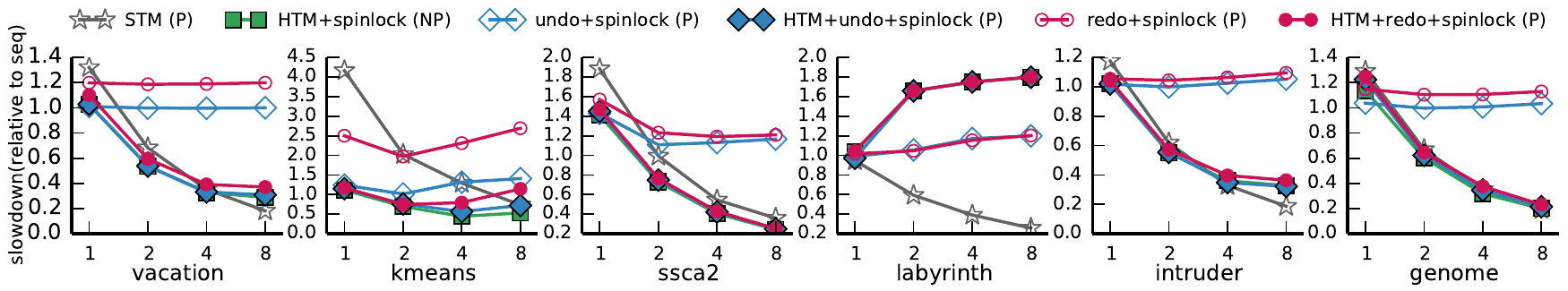}
	\caption{\small TSX-enabled hardware with real NVM and emulated persistent caches by number of threads (X axis).
	(P) crash-consistent  (NP) not crash-consistent.} 
	\label{fig:Mthr-real-pcache} 
		 \end{figure*}

\subsection{Persistent CPU caches}
\label{sec:pcache}

We use our bare-metal test-bed to emulate persistent caches. TSX ensures \cs. We show the results in Figure~\ref{fig:Mthr-real-pcache}.
As expected, \undospinlock and \redospinlock perform the worst and exhibit poor scalability because they serialize all transactions using a global spinlock.
\tsxundospinlock (\tsxredospinlock) is $2.4\times$ ($2.4\times$) faster than \undospinlock (\redospinlock) on average for all workloads at 4 threads.  
The only exception is labyrinth, which causes frequent transaction aborts due to overflows. 
%
%However, when the contention increases, transactions start to abort more frequently and the fallback path is taken more often, hurting performance by XX (XX) for <which benchmarks> at 4 threads. 
%	Furthermore, the fallback path uses a global lock that increases false contention.
%	For most applications, this never happens up to 8 threads, with the exceptions of Labyrinth and Genome, which causes frequent aborts.
	While \stm incurs high overheads in low contention scenarios (1 or 2 threads) or when transactions are small, it exhibits good scalability due to its fine-grained locking and generally performs the best at 8 threads and for large transactions. \stm is $1.6\times$ ($2.0\times$), $6.9\times$ ($6.9\times$) faster than \tsxundospinlock (\tsxredospinlock) for vacation and  labyrinth at 8 threads.

 % !TEX root = paper.tex

\section{Evaluating crash-consistency}
\label{sec:st}

In this section, we seek to understand what is the cost of \cc for single-thread applications, i.e., when applications do not 
require synchronization. To do so, we perform an exhaustive study using multiple hardware platforms, using simulation and 
emulation when the actual hardware is not available. 
 We evaluate both transient (\S\ref{sec:st-transient})  and persistent caches (\S\ref{sec:st-persistent}).
We present the results relative to a sequential execution baseline (\textbf{\seq}) that does not provide \cc nor \sync.
We evaluate the \ccms described in Table~\ref{table:tl-impl}.

The goal of this evaluation is to understand the cost of \csty relative to only providing \cc, and whether providing \csty provides 
any benefits from a performance perspective compared to simply providing \cc. 
We see that \csty is a useful implementation property, as it can lower the cost of \cc by scaling applications to multiple threads. 
For example, in the vacation benchmark 
achieving crash-sync-safety for 8 threads improves performance by $0.7\times$ compared to non-crash consistent single-thread execution,
despite the crash-sync-safety property due to ccSTMs scalability.
In contrast, the undo log causes a slowdown of $0.85\times$ to achieve crash consistency only (for a single-thread execution).
In this section, we breakdown the costs of crash-consistency and characterize single-thread applications.

\noindent\textbf{Summary of crash-consistency results.}
We find that the persistence domain plays a crucial role in choosing the best \cc method.
 The same mechanisms have very different behavior and performance characteristics on systems with transient caches versus systems with persistent caches. 
In particular, \htm is an interesting case-study. 
For systems with persistent caches, \htm guarantees \cc out of the box, with no architectural changes, while for transient caches, \htm needs architectural changes to ensure \cc. 
In both cases, \htm also provides correct \sync, and incurs the associated costs, although all applications we consider in this section are single-threaded and do not require \sync. 

From our empirical results, we can draw the following conclusions:
(1a) In systems with transient caches, \htm benefits \cc, despite its \sync costs and required architectural changes. The reason for this is that \htm based \cc systems are able to reduce the number of expensive cache line flush and fence instructions used in pure software \undo/\redo logging techniques.
(1b) The choice between \undo and \redo logging vastly depends on the application characteristics and the size of the read and write sets of the transactions. 
(2a) In systems with persistent caches, the \htm benefit for \cc is reduced, as software logging mechanisms do not require expensive flush and fence instructions anymore. In this case, 
the \htm's \sync overheads become apparent.
%, even though its implementation is more lightweight than for transient caches, with no additional hardware mechanisms required for \cc. 
(2b) \undo logging is the best choice for ensuring \cc when caches are persistent, since \redo logs still suffer from read-indirection overheads.

Overall, persistent caches provide a significant advantage, as the overhead of \cc on average over \seq is only 1\% for the best method (\undo), compared to 6\% for best method when caches are transient (\htmredo).

%%%%% 
% Note: what would a ccHTM for a single thread look like if it provides persistence but not synchronization? 

\subsection{Transient CPU caches}
\label{sec:st-transient}

We compare the performance of the various \ccms on systems with transient CPU caches (\autoref{ptcache}),
%where data has to be explicitly written back from the caches to be considered persistent. 
%\ic{ref background transient caches section}
%We perform this comparison 
on both real hardware and our architectural simulator (\autoref{sec:methodology}). 
%\ic{ref Methodology section}

%\begin{figure}[tbp] 
%	\centering
%	\includegraphics[width=\linewidth]{plot/singleallcompressed-transientplus.pdf} 
%	\caption{\small (Broadwell) TSX-enabled hardware, transient caches with 
%	 transaction success rate for methods using HTM. We truncate large bars in Kmeans (values in red). (P) crash-consistent; (P*) it approximates a crash-consistent solution; (NP) not crash-consistent. } 
%	 \label{fig:1thr-real} 
%\end{figure}

%\begin{figure}[tbp] 
%	\centering
%	\includegraphics[width=\linewidth]{plot/singleallcompressed-transientaep.pdf} 
%	\caption{\small (Skylake) TSX-enabled hardware, transient caches with 
%	 transaction success rate for methods using HTM. We truncate large bars in Kmeans (values in red). (P) crash-consistent; (P*) it approximates a crash-consistent solution; (NP) not crash-consistent. } 
%	 \label{fig:1thr-real} 
%\end{figure}

\begin{figure}[tbp] 
	\centering
	\includegraphics[width=\linewidth]{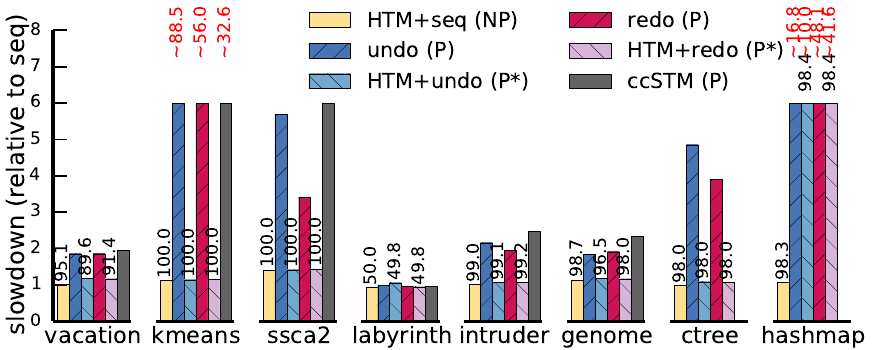} 
	\caption{\small TSX-enabled hardware, with real NVM and transient caches. We show 
	 transaction success rate for methods using HTM (values in black). We truncate large bars in Kmeans and hashmap (values in red). 
	 (P) crash-consistent; (P*) approximates a crash-consistent solution; (NP) not crash-consistent. } 
	 \label{fig:1thr-real} 
\end{figure}

\begin{figure}[tbp]   
	\centering
	\includegraphics[width=\linewidth]{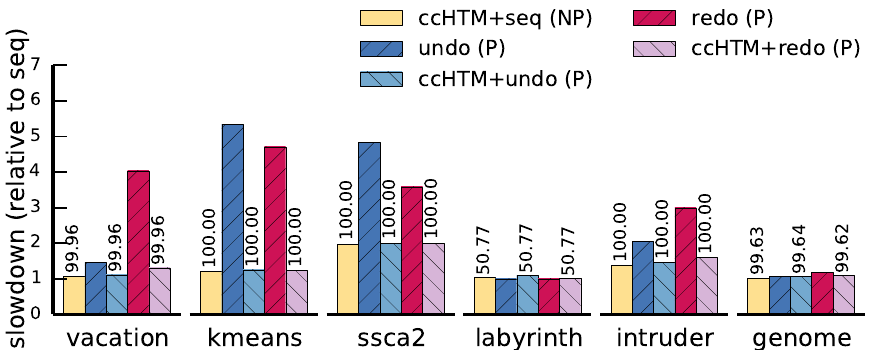} 
	\caption{\small Simulation, transient caches. We show transaction success rate on top of the methods using HTM. For each method, we specify 
	 if it is crash-consistent (P) or not (NP). } 
	\label{fig:1thr-sim} 
\end{figure}

\noindent{\textbf{Real.} }
%\subsubsection{Real.} 
In this experiment, we evaluate all \ccms on the real hardware using a server with support for TSX 
%to approximate the performance of a real \cchtm. In this case, the \htm and \stm provide 
%only \sync, but no \cc. We use \htm with \undo and \redo fallback (\tsxundo and \tsxredo) to approximate 
%\cchtm and we implement \ccstm. 
and we show the results in Figure~\ref{fig:1thr-real}.
As expected, all \ccms increase execution time compared to a non-crash-consistent baseline (\seq).
However, \tsxundo and \tsxredo significantly outperform their pure software counterparts (\undo and \redo), improving performance by
as much as $98\times$ and $97\times$, respectively. This is due to the \htm reducing the number of fences and read-indirection 
for the transactions that succeed. 
The only exception is Labyrinth, where \tsxundo and \tsxredo perform similar to their software counterpart, 
due to more frequent aborts caused by large transaction sizes (Labyrinth). 
%While the Hashmap's small
%transactions are HTM friendly, the overall execution time is dominated by the hash-rebuild (large transaction) that also causes capacity aborts. -- no problem with hashmap
%\ic{what's the reason for Hashmap? What's transaction success rate?}
%or unsupported instructions (Genome). 
These results indicate the best performance that we can expect from a \cchtm on real hardware, 
as we approximate the performance using HTM+undo/redo that only provides \cc on the fallback path, but not inside the hardware transaction.
%as we only evaluate HTM+undo/redo, which only provides \cc on the fallback path, but not inside a hardware transaction. 
Therefore, these results measure the upper limit of ccHTM. For a more conservative estimate of ccHTM performance, we also evaluate it in the simulator (Fig.~\ref{fig:1thr-sim}).
The transaction success rate\footnote{TSX transactions are best-effort, so they might abort even for single-thread 
workloads, when there are no conflicts.}
 varies from from 49\% (Labyrinth) to 100\% (Kmeans) - also 
shown in Figure~\ref{fig:1thr-real}.

\tsxundo and \tsxredo are on average 14\% and 12\%, respectively, of the ideal baseline, \seq,
showing that  
\cc can be ensured at a small performance penalty using \htm.
Although not needed for single-threaded applications, \htm also provides \sync. To understand this additional overhead, we also evaluated \htmseq, which ensures \sync when transactions succeed, but not on the fallback path. 
\htmseq, incurs overheads, ranging from 2.5\% (intruder) to 13\% (genome), even when no crash consistency guarantees are provided. These overheads are due to hardware book-keeping to execute transactions and transaction aborts. 
%\ak{It is a little weird that HTM+seq is slower than HTM+undo/redo in some cases (genome).} 
% I checked the results files.The lowest number of HTM+seq is better than HTM+undo/redo. Just that the final number depend on middle 3 average.
% HTM+redo had a good 3 runs from the results. Hence it is marginally beating HTM-seq. The HTM success rate is high in this benchmark. 
%\ic{Also need to compare HTM+undo/redo with HTM+seq}

All pure software \ccms (\undo, \redo and \ccstm) greatly increase execution times --
as much as $41\times$, $89\times$ and $33\times$ respectively (for Kmeans).
\redo logging generally performs better or as well as \undo logging for the workloads evaluated, but the 
results highly depend on the number of reads and writes in the transaction. 
%The two mechanisms have different overheads, 
\undo suffers the overhead of flushing and fences for every write, while \redo suffers the overhead of read indirection, proportional 
to the number of reads and the size of the log. 
%
%However, the overhead of the \redo log depends on the size of the log
%
% for 
%workloads with many reads, but only if there are many writes also (the overhead of read indirection depends on the size of the log).
\ccstm incurs the highest overhead across all workloads. We 
attribute this to the higher \sync overheads of the \ccstm, in addition to the software logging overheads like \undo and \redo.

\noindent{\textbf{Simulated.}}
%\subsubsection{Simulated.}
In this experiment, we use the simulator to evaluate 
\cchtm for transient caches with the proper architectural changes. 
Figure~\ref{fig:1thr-sim} shows the results.
The trends from the real hardware still largely hold here.
\sysundo and \sysredo have lower overheads than their software counterparts by as much as  $3.2\times$ and $2.7\times$ (Kmeans) respectively.
And they both come within $1.2\times$ of the ideal baseline, \seq.
Even with full support for \cc, \cchtm outperforms other methods.
The only exception is Labyrinth, where \cchtm suffers comparable overheads to its software counterparts due to frequent transaction aborts.
However, the transaction abort rate is less in the simulator than on real hardware as the simulator models overflow and some unsupported instructions, but not all events that would cause a transaction to abort on real hardware. 
We attribute these differences to inaccuracies between the hardware implementation details within the simulator and proprietary commercial hardware.
\sysseq adds 28.62\% overhead on average for all benchmarks compared to \seq, highlighting 
that the architectural changes made to the \htm have fairly low impact. 

\subsection{Persistent CPU caches}
\label{sec:st-persistent}

\begin{figure}[t] 
	  \centering
		  \includegraphics[width=\linewidth]{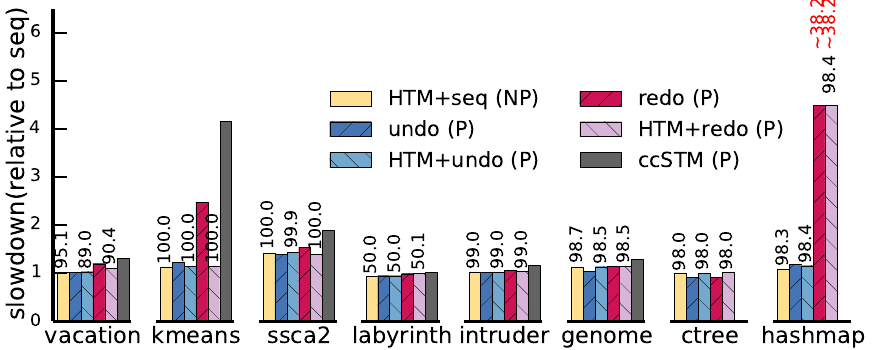} 
			\caption{\small TSX-enabled hardware with real NVM, emulating persistent caches. 
				We show transaction success for methods using HTM (values in black). We truncate large bars in hashmap (values in red).
			(P) crash-consistent; (NP) not crash-consistent. }
			\label{fig:1thr-pcache} 
		\end{figure}

In this experiment, 
%we consider platforms where the persistence domain includes the 
%cache hierarchy .  
%\ic{ref background on persistent caches}
we use our bare-metal testbed 
(\autoref{sec:methodology}) to emulate persistent caches (\autoref{ptcache}). 
We show the results in Figure~\ref{fig:1thr-pcache}.
Unlike for transient caches, here \tsxundo and \tsxredo perform worse than \undo and \redo by at most 10\%.
% and 10\%, respectively. 
When caches are persistent, \undo and \redo logging techniques no longer have to use expensive 
cache line flush and fence instructions to ensure data consistency, while the \htm still has the overhead of 
ensuring \sync. To quantify the overhead of the \htm, we measure \tsxseq, which is up to 40\% slower than \seq, while \undo and \redo are up to 40\% and $37\times$ slower. The \stm has even higher \sync overhead on average, being up to $3.1\times$ slower than \seq. 
Moreover, \undo and \redo perform similarly in most cases, except in the case of hashmap and Kmeans, where \undo performs better than \redo. Although both methods are faster because they don't require fences and flushes, \redo still has the overhead of read indirection in certain workloads with many reads and writes. Moreover, an application can be tuned to use one technique or the other, which we show with hashmap as an example. Hashmap is tuned to use undo logging with PMDK library and thus perform better with undo logging.

\section{Related Work}

%Our work builds on two active areas of research: transactional memory and NVM crash-consistency.
%First work to propose using them together were Mnemosyne~\cite{mnemosyne} and NV-Heaps~\cite{nvheaps}.
Mnemosyne~\cite{mnemosyne} and NV-Heaps\cite{nvheaps} were the first to extend an STM for providing 
persistence when caches are transient. Recent research~\cite{pisces,timestone} further 
improves the scalability of durable STMs with better
concurrency-control protocols,  
DRAM+NVM hybrid logging-schemes, etc.  
The ccSTM that we evaluate closely follows the design proposed by
Mnemosyne, but the insights are 
equally applicable to other durable STM proposals.
%
%persistent STM and provided
%persistency at word granularity and object granularity respectively. 
%
Atlas~\cite{atlas} extends critical sections based on locks with persistence semantics and 
gurantees failure-atomicity of outer-most critical
sections. NVThreads~\cite{nvthreads} builds on 
Atlas to provide a drop-in replacement for pthreads that enables NVM crash-consistency.
SFR~\cite{sfr} on the other hand, provides persistence at thread regions delimited by 
synchronization operations. 
Atlas~\cite{atlas}, SFR~\cite{sfr} and other proposals~\cite{kollilanguage,ido} essentially
use data-race-free (DRF) property of correctly synchronized programs and support compiler/ISA
level fast-persistence with NVM. Therefore, our lock based redo/undo log evaluations broadly
model the performance characteristics of these proposals.

%Unlike our lock-based code, Atlas maintains the semantics provided by the lock implementation and does not adopt the transactional 
%semantics. This difference becomes apparent especially for nested critical sections. 
%NVThreads~\cite{nvthreads} provides a drop-in replacement for pthreads that enables NVM crash-consistency.

%
%It uses compiler support to instrument code within transactional blocks.
%NV-Heaps~\cite{nvheaps} in contrast provides a object based programming model and thus maintains consistency at the
%object granularity.
%
%Atlas~\cite{atlas} provided durability support for lock based applications thus making them automatically
%crash-consistent. Atlas instruments the code enclosed by
%locking and support the crash consistent updates by means of logging. NVThreads~\cite{nvthreads} uses principals
%developed in Atlas to build a drop-in replacement for pthreads that enable NVM main memory ready
%applications.
%
%Unfortunately, purely STM-based solution can provide poor performance as evident from the evaluation section.
%The first work to propose HTM for persistence have been DudeTM~\cite{dudetm}.
DudeTM~\cite{dudetm} extends an HTM for persistence using 
%solves the problem of hardware controlled write ordering within a transaction altogether by
%creating 
a shadow copy of the NVM data in volatile memory. Unlike DudeTM, 
\sys does not incur the overhead of additional shadow copies. 
%It maintains
%NVM data structure consistency by 1) maintaining a redo log in persistent memory for every update
%2) updating the original NVM data structures using the persisted log. In a way dudeTM avoids the
%read-redirection problem inherent to redo logging at the cost of additional data copy.
%
%The fallback path when transactions fail in hardware is important for an efficient solution.
PHTM~\cite{phtm} extends a persistent HTM using non transactional stores
and transparent flush semantics to ensure crash-consistency.
%Their solution was limited in hardware support and required marker
%tables to avoid concurrent NVM-logging of variables whereas in our solution
%the same is implicitly supported by HTM read/write set. 
PHTM was extended in PHyTM~\cite{phytm} by adding an STM in the fallback path. 
%Instead of global lock in the fallback path it uses two phase locking at the word granularity in the
%fallback path. Similar to PHTM the transactions are accelerated
%using persistent HTM blocks. %The proposal introduces PHTM and PSTM integration algorithm
%in the form of fast HTM path, slow HTM path and STM path. 
Both PHTM and PHyTM emulate
logging inside the HTM region using regular load/stores instead of their non transactional stores
and transparent flush support.
Thus, their design affects the read/write set and capacity
aborts. In addition, PHTM and PHyTM provide only an approximation of their system performance using 
a TSX host, but no implementation of their proposed hardware extensions. NV-HTM~\cite{nvhtm} 
introduces HTM accelerated persistent memory transactions without changing the existing HTM hardware
protocols. NV-HTM differs durable log-commit till HTM-end for correctness reasons, and thus misses
out on overlapped durable log-writes(\cchtm).

%We provide a realistic evaluation with hardware changes in a simulator.

%Hardware support for NVM is being proposed in industry and academia. 
%for efficient support of NVMs. 
Intel added new instructions \texttt{clflushopt} and \texttt{clwb} for efficient
 transient cache-line flush~\cite{intel-flush}. %Using these semantics and extending them,
%various researchers have proposed 
Researchers have proposed persistency models~\cite{whisper,Pelley:2014,Kolli:2016,Lu:2014,Alshboul:isca:2018} to reason about crash-consistency for NVM.
Various proposal to perform efficient logging and paging for NVMs are also proposed~\cite{Ogleari:2018,Seo:2017,Huang:2015,Wang:2014,Joshi:2017,Shin:2017}. 
Additionally, there has been increasing interest in developing persistent transactional memories 
or memory architectures for NVMs with transient caches~\cite{Giles:ismm:2017,Giles:spaa:2017,Wang:cal:2015,Joshi:isca:2018,Doshi:hpca:2016,Kolli:asplos:2016,Memaripour:eurosys:2017,phtm,Hassan:pact:2015,Dulloor:eurosys:2014}.
Moreover, various applications have been ported to use NVM and have been shown to increase performance~\cite{Marathe:2017,Zhang:msst:2015}.
%More recently, HOPS~\cite{whisper} proposed new hardware ISA primitives to support both ordering
%and durability independently -- namely OFENCE and DFENCE.
%Using them, they offload the persistent updates to memory controller by augmenting
%the persistent stores with ordering points and durability points.

Persistent caches~\cite{Narayanan:2012,Wang:2014,Zhao:2013} may become prevalent. 
% which we study as part of our work.
Recent work %showed 
uses logging with persistent caches to provide durability guarantees~\cite{Izraelevitz:2016, Marathe:2018}. 
Our work studies tradeoffs of providing \cs for both persistent and transient caches.

\section{Conclusion}

%Non-volatile memory technologies promise fast, byte-addressable persistent storage.
%However, programming for NVMs is complex and error-prone. 
%In this paper, we explored the transactional programming model for both crash consistency and synchronization simultaneously, by combining them in the same runtime. 

In this paper, we provide an extensive study of NVM crash-consistency mechanisms for persistent and transient caches. 
Our findings indicate that the persistence domain determines the cost of crash-consistency, but we can reduce the overhead by scaling-up to multiple threads and combining crash-consistency with synchronization.

%\balance
%%%%%%% -- PAPER CONTENT ENDS -- %%%%%%%%

%%%%%%%%% -- BIB STYLE AND FILE -- %%%%%%%%
\bibliographystyle{IEEEtran}
\bibliography{refs}
%%%%%%%%%%%%%%%%%%%%%%%%%%%%%%%%%%%%

\end{document}